\newif\ifsupp
\supptrue 

\documentclass[superscriptaddress,twocolumn,10pt,prl,floatfix,nofootinbib,aps]{revtex4-2}

\input{glyphtounicode}
\usepackage[T1]{fontenc}
\usepackage[utf8]{inputenc}
\usepackage[english]{babel}
\usepackage{amsmath} 
\usepackage{amsfonts} 
\usepackage{amssymb}
\usepackage{graphicx} 
\usepackage{hyperref}
\hypersetup{bookmarksopen,bookmarksnumbered,
                colorlinks,
                linkcolor=blue,
                citecolor=blue}
\usepackage{natbib}
\setcitestyle{super}
\usepackage{array}
\usepackage[table]{xcolor}
\usepackage{xcolor}
\usepackage{xr}
\usepackage{braket}
\usepackage{booktabs}
\usepackage{multirow}
\usepackage{dsfont} 
\usepackage{caption}

\usepackage{tabularx}
\usepackage{siunitx}
\DeclareSIUnit\permille{\text{\textperthousand}}

\usepackage{subcaption}
\usepackage{bigstrut}
\usepackage{svg}
\usepackage{upgreek}

\usepackage{newfloat}

\DeclareFloatingEnvironment[name={\textbf{Supplementary Figure}}]{suppfigure}

\newcommand{\aref}[1]{\hyperref[#1]{Appendix~\ref*{#1}}}
\DeclareCaptionLabelSeparator{line}{\hspace{2pt}\bf{|}\hspace{2pt}}

\begin{document}

\captionsetup[table]{name={\bf{Table}},labelsep=period,justification=raggedright,font=small,singlelinecheck=false}
\captionsetup[figure]{name={\bf{Figure}},labelsep=line,justification=raggedright,font=small,singlelinecheck=false}

\renewcommand{\equationautorefname}{Eq.}
\renewcommand{\figureautorefname}{Fig.}
\renewcommand*{\sectionautorefname}{Sec.}

\title{Optimal operating temperature for industry-compatible silicon spin quantum computing: colder is not necessarily better}

\author{\textbf{Paul Steinacker}}
\email{paul.s@diraq.com}
\affiliation{Diraq, Sydney, New South Wales, Australia}
\affiliation{School of Electrical Engineering and Telecommunications, University of New South Wales, Sydney, New South Wales, Australia}
\author{Amanda E. Seedhouse}
\affiliation{Diraq, Sydney, New South Wales, Australia}
\affiliation{Universit\'e Grenoble Alpes, CNRS, LPMMC, 38000 Grenoble, France.}
\author{Nard Dumoulin Stuyck}
\author{Tuomo Tanttu}
\author{MengKe Feng}
\author{Santiago Serrano}
\author{Ensar Vahapoglu}
\author{Samuel K. Bartee}
\author{Philip Y. Mai}
\author{Alexis Shaw}
\affiliation{Diraq, Sydney, New South Wales, Australia}
\affiliation{School of Electrical Engineering and Telecommunications, University of New South Wales, Sydney, New South Wales, Australia}
\author{Andreas Nickl}
\affiliation{School of Electrical Engineering and Telecommunications, University of New South Wales, Sydney, New South Wales, Australia}
\author{Sebastian Pauka}
\author{Brendan Harlech-Jones}
\author{Juan P. Dehollain}
\affiliation{ARC Centre of Excellence for Engineered Quantum Systems, School of Physics, University of Sydney, Sydney, New South Wales, Australia}
\affiliation{Emergence Quantum, Sydney, New South Wales, Australia}
\author{Fay E. Hudson}
\author{Kok Wai Chan}
\affiliation{Diraq, Sydney, New South Wales, Australia}
\affiliation{School of Electrical Engineering and Telecommunications, University of New South Wales, Sydney, New South Wales, Australia}
\author{Thomas A. Ohki}
\author{David Reilly}
\affiliation{ARC Centre of Excellence for Engineered Quantum Systems, School of Physics, University of Sydney, Sydney, New South Wales, Australia}
\affiliation{Emergence Quantum, Sydney, New South Wales, Australia}
\author{Christopher C. Escott}
\affiliation{Diraq, Sydney, New South Wales, Australia}
\author{Chih Hwan Yang}
\author{Wee Han Lim}
\author{Arne Laucht}
\affiliation{Diraq, Sydney, New South Wales, Australia}
\affiliation{School of Electrical Engineering and Telecommunications, University of New South Wales, Sydney, New South Wales, Australia}
\author{Andre Saraiva}
\email{andre@diraq.com}
\affiliation{Diraq, Sydney, New South Wales, Australia}
\author{Andrew S. Dzurak}
\affiliation{Diraq, Sydney, New South Wales, Australia}
\affiliation{School of Electrical Engineering and Telecommunications, University of New South Wales, Sydney, New South Wales, Australia}
\author{Jared H. Cole}
\email{jared.cole@diraq.com}
\affiliation{Diraq, Sydney, New South Wales, Australia}
\affiliation{School of Science, RMIT University, Melbourne, Victoria, Australia}
\date{\today}

\begin{abstract}
Silicon spin qubits are a leading candidate for large-scale quantum computing owing to their compatibility with semiconductor manufacturing. However, scaling to useful fault-tolerant processors will likely generate thermal loads that exceed the cooling power available at millikelvin temperatures. Raising the operating temperature eases cooling requirements but reduces gate fidelity, increasing the overhead of quantum error correction. Identifying the operating temperature that minimizes total power consumption is therefore a key challenge for commercially viable quantum computers.
Here, we use gate set tomography to benchmark two-qubit silicon chips fabricated in both industrial and academic environments over a range of temperatures. Elevated temperatures substantially shorten coherence times and increase gate and state-preparation-and-measurement infidelities. Based on these measurements, we develop a general power model for silicon quantum computers that combines cryogenic cooling requirements with error-correction overheads. We show that a finite optimal operating temperature exists and is strongly influenced by a crossover temperature near \SI{1}{\kelvin} in current devices, above which gate fidelity degrades rapidly. These results connect device-level fidelity limitations to system-level power requirements, providing design guidelines for large-scale silicon quantum computers.
\end{abstract}

\maketitle
\noindent
Operation of a utility-scale quantum computer will require millions of physical qubits operating at error rates well below the threshold of the underlying quantum error-correction code. Achieving such performance incurs substantial energy costs across quantum-computing platforms. Many leading architectures rely on cryogenic cooling of either the qubits~\cite{veldhorst_addressable_2014,veldhorst_two-qubit_2015,vandersypen_interfacing_2017,yoneda_quantum-dot_2018,noiri_fast_2022,takeda_quantum_2022,madzik_precision_2022,members_of_the_hrl_quantum_team_digitally_2026,devoret_superconducting_2013,barends_superconducting_2014,arute_quantum_2019,krinner_realizing_2022,kim_evidence_2023,google_quantum_ai_suppressing_2023,google_quantum_ai_and_collaborators_quantum_2025} or the measurement systems~\cite{zhong_quantum_2020,madsen_quantum_2022,cheng_100-pixel_2023,bartolucci_fusion-based_2023}, whereas others require power-intensive infrastructure including vacuum systems and high-performance laser control~\cite{lekitsch_blueprint_2017,bruzewicz_trapped-ion_2019,mehta_integrated_2020,romaszko_engineering_2020,srinivas_high-fidelity_2021,pino_demonstration_2021}. As demonstrated by the growth of conventional data centers~\cite{de_vries_growing_2023,luccioni_power_2024,chen_how_2025}, once computing systems reach sufficient scale, energy becomes the dominant operating cost and ultimately determines their commercial viability. Although current noisy intermediate-scale quantum (NISQ) devices~\cite{preskill_quantum_2018} are still largely operated as laboratory experiments, utility-scale quantum computers will require a whole-system approach that jointly optimizes computational performance and energy consumption.

\begin{figure} 
    \includegraphics[width = 1\linewidth]{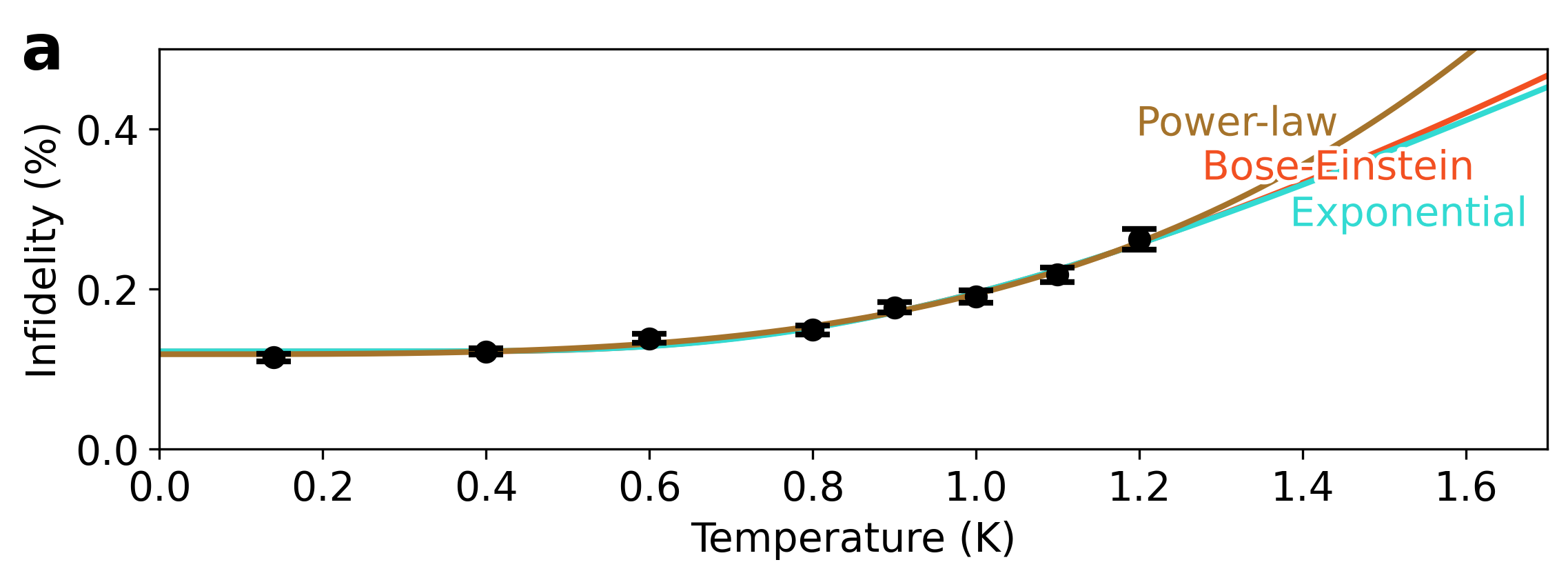}
    \includegraphics[width = 1\linewidth]{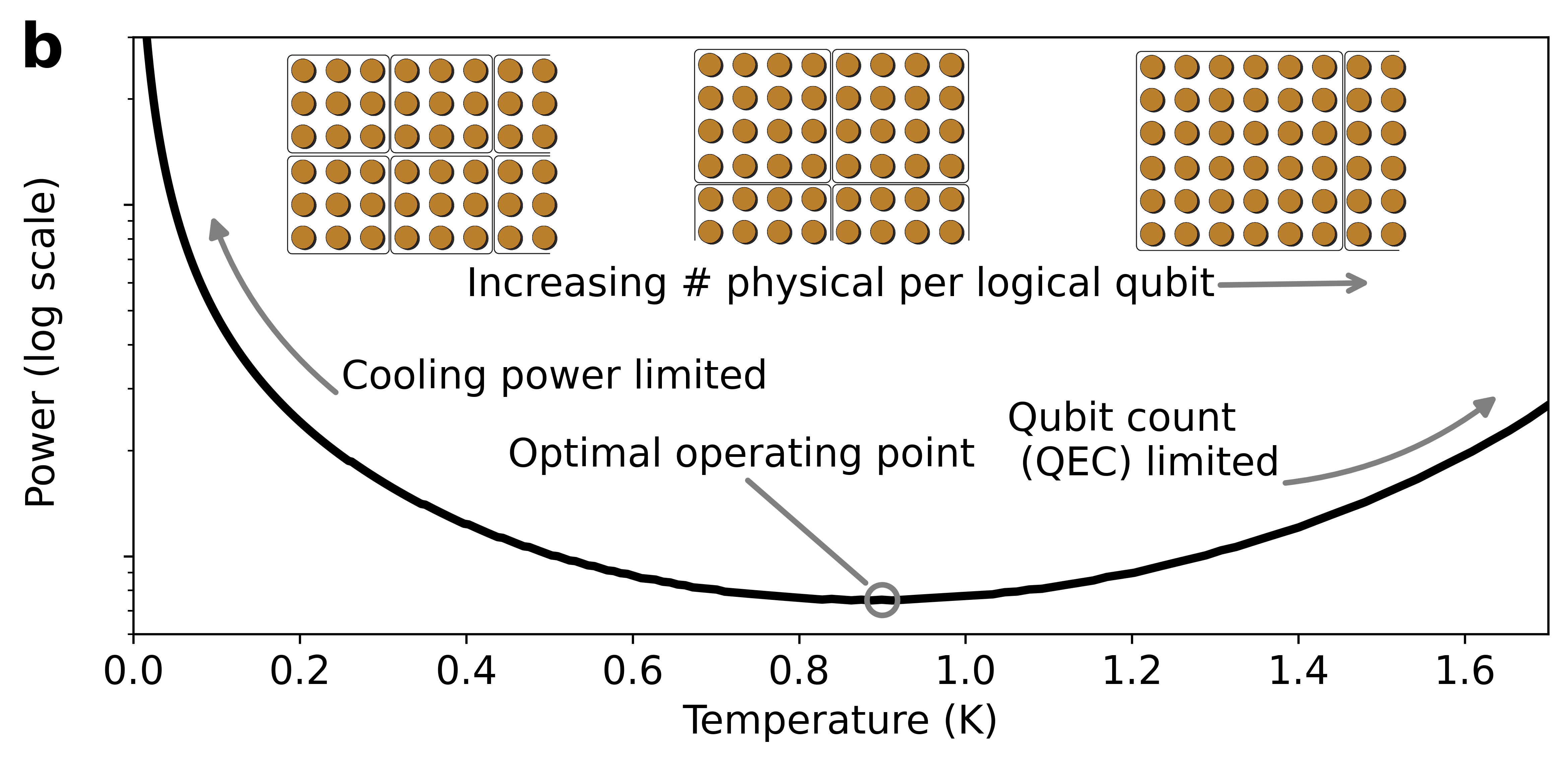}
    \caption{\textbf{Fidelity and operating power as a function of temperature.} 
    \textbf{a,} Illustration of qubit operation infidelity examples as a function of temperature. The black dots represent Clifford gate infidelity measured with RB taken from ref.~\cite{huang_high-fidelity_2024}. Error bars represent the \SI{95}{\percent} confidence level. The temperature response is fitted to three examples of functional forms motivated by common quantum computing modalities.   
    \textbf{b,} Operating power as a function of temperature, illustrating the trade-off between high cryogenic cooling power consumption, and high qubit count or reduced logical qubit count for fixed hardware.}
    \label{fig:main_fig0}
\end{figure}

Achieving commercially relevant quantum computers with silicon qubits requires all qubits and their control electronics to be integrated into a single package operating at cryogenic temperatures~\cite{members_of_the_hrl_quantum_team_digitally_2026}. The high electronic component density required to achieve this will most likely generate a thermal load larger than the cooling power available in today's most powerful dilution refrigerators at millikelvin temperatures~\cite{krinner_engineering_2019,franke_rents_2019}. More generally, cryogenic systems are inherently high-power consumers, as Carnot’s theorem~\cite{carnot_reflexions_1986} dictates that the minimum power required to cool a system increases drastically with decreasing operating temperature. In practice, real cryogenic systems operate far below the Carnot efficiency, with the available cooling power increasing by approximately five orders of magnitude when the operating temperature is raised from $\SI{20}{\milli\kelvin}$ to \SI{1}{\kelvin} for comparable electrical input power~\cite{pobell_matter_2007,uhlig_dry_2015,gonzalez-zalba_scaling_2021}. Raising the operating temperature substantially relaxes refrigeration requirements, but comes at the expense of qubit performance. Higher temperatures generally degrade gate and measurement fidelity, although the specific dependence varies between qubit modalities (Fig.~\ref{fig:main_fig0}a). Within a fault-tolerant architecture, higher physical error rates require larger quantum error-correction overheads, increasing the number of physical qubits and control resources needed to realize a fixed number of logical qubits~\cite{fowler_surface_2012,beverland_assessing_2022}. Operating temperature therefore determines total system power through competing refrigeration and fault-tolerance costs, suggesting that a finite power-optimal operating temperature should exist (Fig.~\ref{fig:main_fig0}b).

Operation above the millikelvin regime increases the available thermal budget, potentially enabling the integration of cryo-CMOS control electronics~\cite{pauka_cryogenic_2021,xue_cmos-based_2021,ruffino_cryo-cmos_2021,bartee_spin-qubit_2025}. However, higher temperatures generally degrade qubit fidelity, increasing the resources required for fault-tolerant quantum computation. Quantifying this trade-off between refrigeration requirements and error-correction overhead has therefore emerged as an important architectural challenge for cryogenic quantum-computing platforms~\cite{auffeves_quantum_2022,fellous-asiani_optimizing_2023}. Yet no framework currently links experimentally measured temperature-dependent qubit performance to the total power consumption of a fault-tolerant quantum computer, leaving the energy-optimal operating temperature unknown.

\section{Temperature dependence of qubit performance}
\begin{figure*}[bth]
    \includegraphics[width = 1\linewidth]{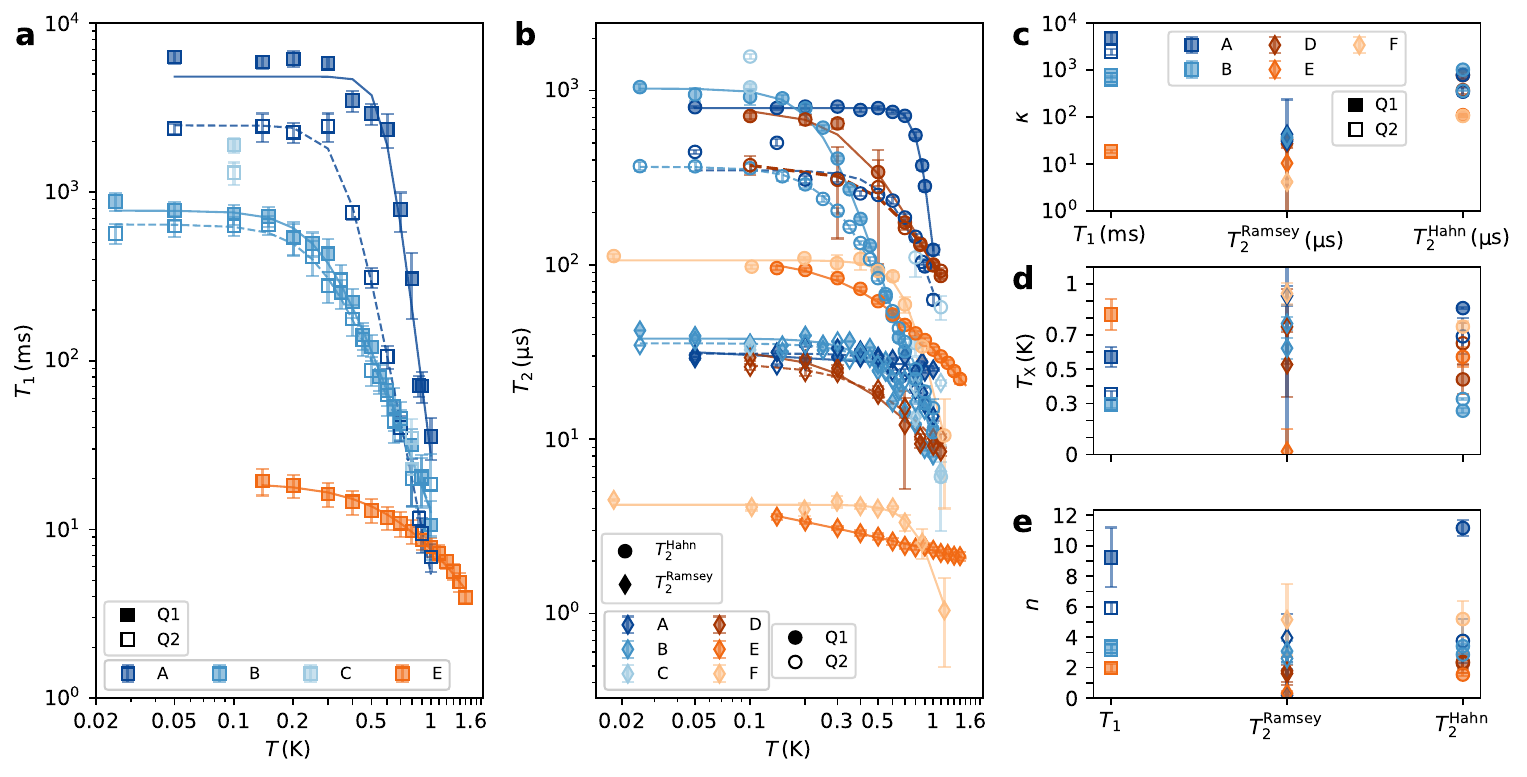}
    \caption{\textbf{Relaxation and coherence time.} 
    \textbf{a,} Relaxation time, $T_{1}$, of devices A--C and E as a function of temperature for Q1 and Q2. 
    \textbf{b,} Dephasing times, $T_{2}^{\mathrm{Ramsey}}$ and $T_{2}^{\mathrm{Hahn}}$, of devices A--F as a function of temperature for Q1 and Q2.
    \textbf{c--e,} Power law fitting parameters $\kappa$, $T_{\mathrm{X}}$, and $n$ for $T_1$, $T_2^{\mathrm{Ramsey}}$, and $T_2^{\mathrm{Hahn}}$ of devices A--F for Q1 and Q2. Device C was excluded from the fit due to the limited amount of data. 
    For comparison, relaxation and coherence times for devices E and F are imported from ref.~\cite{huang_high-fidelity_2024} and ref.~\cite{bartee_spin-qubit_2025}, respectively. Error bars represent the \SI{95}{\percent} confidence level.
    }
    \label{fig:main_fig1}
\end{figure*}

\noindent
In this work, we evaluate the temperature dependence of silicon spin-qubit performance using devices fabricated in both industry-compatible and academic facilities. Three industry-compatible devices (A--C) were fabricated in a pilot $\SI{300}{\milli \meter}$ wafer line at imec, whereas three academic devices (D--F) were fabricated at the Australian National Fabrication Facility (ANFF). Despite differences in fabrication environment and absolute device performance, both device families exhibit a remarkably similar degradation with increasing temperature, suggesting a common underlying physical origin. We find that the measured performance metrics are generally well described by a power-law dependence of the form
\begin{equation}\label{eq:infidelity}
    {\cal{M}} = \kappa \left[ 1 + \left( \frac{T}{T_X} \right)^n \right],
\end{equation}
where $T$ is the operating temperature, $\cal{M}$ is the performance metric, $\kappa$ is the low-temperature infidelity, and $T_X$ is a cross-over temperature above which the power-law dependence with exponent $n$ becomes dominant. As performance metrics, we consider the relaxation rate (1/$T_1$), dephasing rate (1/$T_2$), and qubit operation infidelity ($1-\cal{F}$). Device-to-device variations primarily affect the prefactor $\kappa$, whereas $n$ and $T_X$ capture the temperature dependence that is central to this work. The power-law model provides a conservative description of the observed trends (Fig.~\ref{fig:main_fig0}a), with alternative functional forms yielding qualitatively similar conclusions (see Extended Material).

All silicon metal-oxide-semiconductor (SiMOS) devices were operated as double quantum dots using similar initialization, control and readout protocols. Devices A--C, E and F were operated in the $(P1,P2)=(1,3)$ charge configuration, whereas device D was operated in $(3,3)$. In each case, the unpaired electron in each quantum dot served as a qubit, denoted Q1 and Q2. To isolate temperature-induced changes in performance, each device was optimized at base temperature and subsequently characterized across the full temperature range without further tuning. Measurements of charge-transition broadening as a function of refrigerator temperature were used to estimate the electron temperature. We observe a linear relationship down to approximately \SI{250}{\milli\kelvin}, below which the electron temperature saturates at around \SI{200}{\milli\kelvin}. Additional device and operating details can be found in the Methods section and previous reports on devices A--C~\cite{steinacker_industry-compatible_2025}, D~\cite{steinacker_bell_2025}, E~\cite{huang_high-fidelity_2024} and F~\cite{bartee_spin-qubit_2025}. 

Importantly, devices A--C were fabricated at imec using a $\SI{300}{\milli\meter}$ industry-compatible spin-qubit process optimized for low charge noise and high device uniformity~\cite{stuyck_integrated_2020,li_flexible_2020,stuyck_uniform_2021,elsayed_low_2024}. These devices were fabricated on isotopically enriched silicon with a residual $^{29}$Si concentration of \SI{400}{ppm}. In contrast, devices D--F were fabricated in an academic cleanroom on isotopically enriched silicon substrates with residual $^{29}$Si concentrations of \SI{800}{ppm} (devices D and F) and \SI{50}{ppm} (device E). In these devices, successive Al gate layers were electrically isolated by approximately \SI{4}{\nano\meter} of thermally grown aluminum oxide. Owing to differences in material quality, isotope concentration and fabrication processes, the academic devices are expected to exhibit higher baseline charge noise and, for devices D and F, increased magnetic noise. Furthermore, device F incorporates a cryogenic CMOS controller operating at millikelvin temperatures. Together, these differences provide a diverse testbed for assessing the robustness of temperature-dependent qubit performance across fabrication platforms.

We use gate set tomography (GST)~\cite{blume-kohout_demonstration_2017,nielsen_probing_2020} to benchmark devices A--D and decompose the observed gate errors into their stochastic and Hamiltonian contributions, providing direct insight into the dominant error mechanisms. Among the measured devices, device A exhibits the highest baseline performance, with state-preparation-and-measurement (SPAM), single-qubit gate and two-qubit gate fidelities exceeding \SI{99.9}{\percent}, \SI{99.9}{\percent} and \SI{99}{\percent}, respectively. Furthermore, all measured operation fidelities remain above \SI{99}{\percent} up to temperatures of approximately \SI{0.5}{}--\SI{0.7}{\kelvin}, highlighting the robustness of state-of-the-art silicon spin qubits to operation above the millikelvin regime. Above this temperature, however, performance degrades rapidly, motivating a detailed investigation of the underlying error channels and their implications for large-scale quantum computing.

\subsection{Single-qubit performance}
\begin{figure*}[bth]
    \includegraphics[width = 1\textwidth]{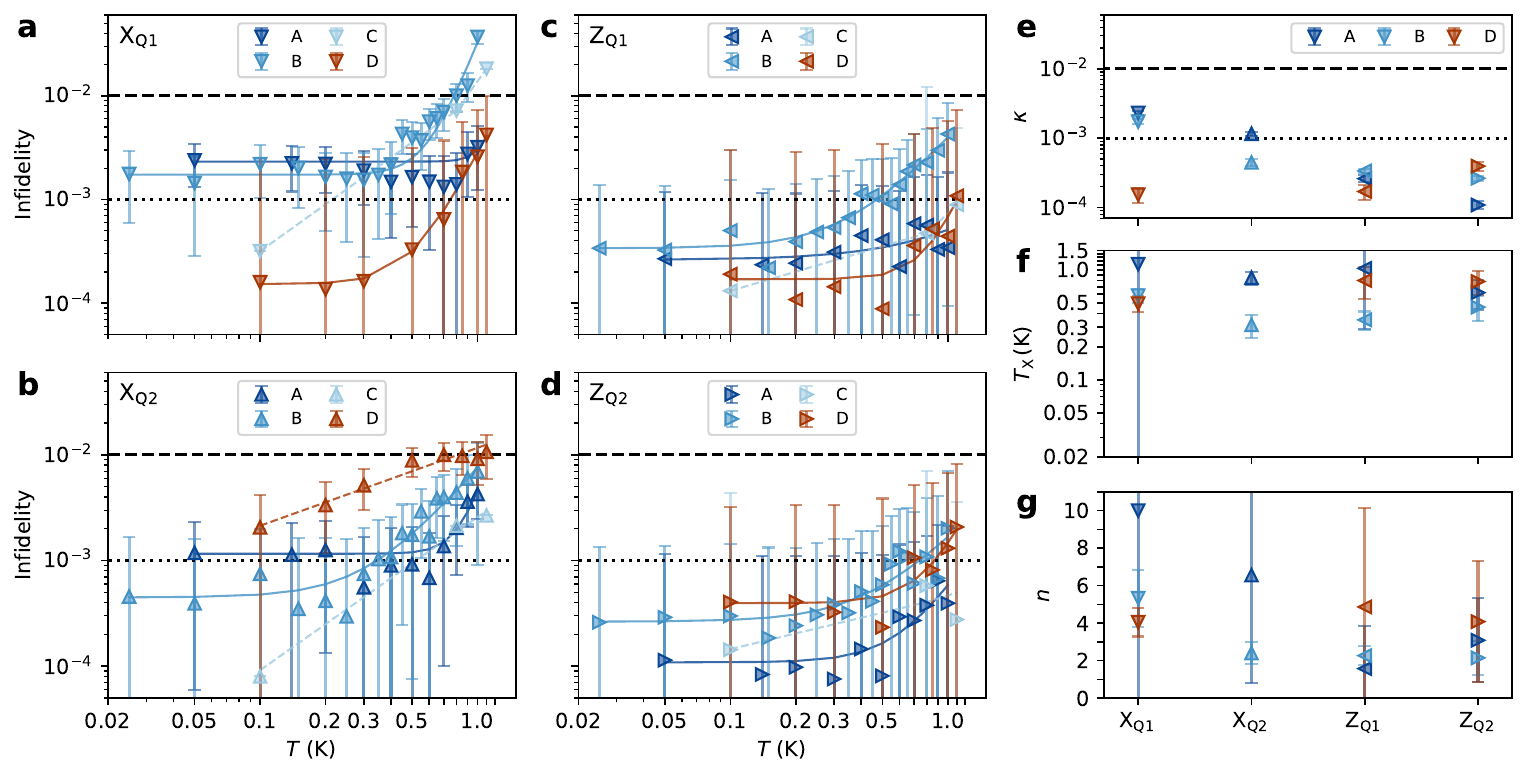}
    \caption{\textbf{Single-qubit infidelity.} 
    \textbf{a--d,} On-target X- and Z-gate infidelity of devices A--D as a function of temperature for Q1 and Q2 extracted from GST. The solid lines are fits to the power law equation~(\ref{eq:infidelity}). The dashed lines are a guide to the eye but no fitting parameters were extracted here.   
    \textbf{e--g,} Power law fitting parameters $\kappa$, $T_{\mathrm{X}}$, and $n$ to the data in \textbf{a}--\textbf{d}. The black dashed (dotted) line represents the \SI{1}{\percent} (\SI{1}{\permille}) infidelity threshold.
    Error bars represent the \SI{95}{\percent} confidence level.
    }
    \label{fig:main_fig2}
\end{figure*}

\noindent
Figure~\ref{fig:main_fig1}a,b shows the spin relaxation and coherence times of devices A--F as a function of temperature. Across all qubits, increasing the operating temperature results in a reduction of $T_1$, $T_2^{*}$, and $T_2^{\mathrm{Hahn}}$ that is well described by the power-law form introduced in Eq.~\ref{eq:infidelity}. The extracted fitting parameters are summarized in Fig.~\ref{fig:main_fig1}c--e. Although the low-temperature performance, quantified by $\kappa$, differs substantially between devices, the crossover temperature and power-law exponent are remarkably consistent. For most qubits, we find $T_\mathrm{X}=\SI{0.3}{}$--\SI{0.8}{\kelvin} and $n=2$--4. This similarity suggests that the elevated-temperature performance is governed by common physical mechanisms despite significant differences in fabrication process, isotopic enrichment and baseline noise levels. We note that both the absolute coherence times and their temperature dependence are expected to depend on the confinement potential and operating voltages. Throughout this study, all devices were operated using a similar tuning strategy optimized for maximum qubit fidelity at base temperature.

Figure~\ref{fig:main_fig2}a--d shows the X- and Z-gate infidelities of devices A--D as a function of temperature, extracted from gate set tomography (GST) experiments performed in the two-qubit operating regime~\cite{madzik_precision_2022}. As observed for the coherence metrics, the gate infidelity increases monotonically with temperature and is well described by Eq.~\ref{eq:infidelity}. The extracted fitting parameters are summarized in Fig.~\ref{fig:main_fig2}e--g. Despite differences in base-temperature performance, with $\kappa$ ranging from approximately $10^{-4}$ to $10^{-3}$, all devices exhibit similar temperature scaling characterized by crossover temperatures of $T_\mathrm{X}=\SI{0.5}{}$--\SI{1}{\kelvin} and power-law exponents of $n=2$--5. The higher crossover temperatures observed for driven operations indicate that coherent control remains relatively robust even as idle-state coherence begins to degrade. This likely reflects the reduced sensitivity of actively driven gates to errors that accumulate during free evolution. Having established the temperature dependence of the single-qubit gates, we next use the full GST results to assess the evolution of two-qubit system including SPAM performance.

\subsection{Two-qubit performance}
\begin{figure*}[bth]
    \includegraphics[width = 1\textwidth]{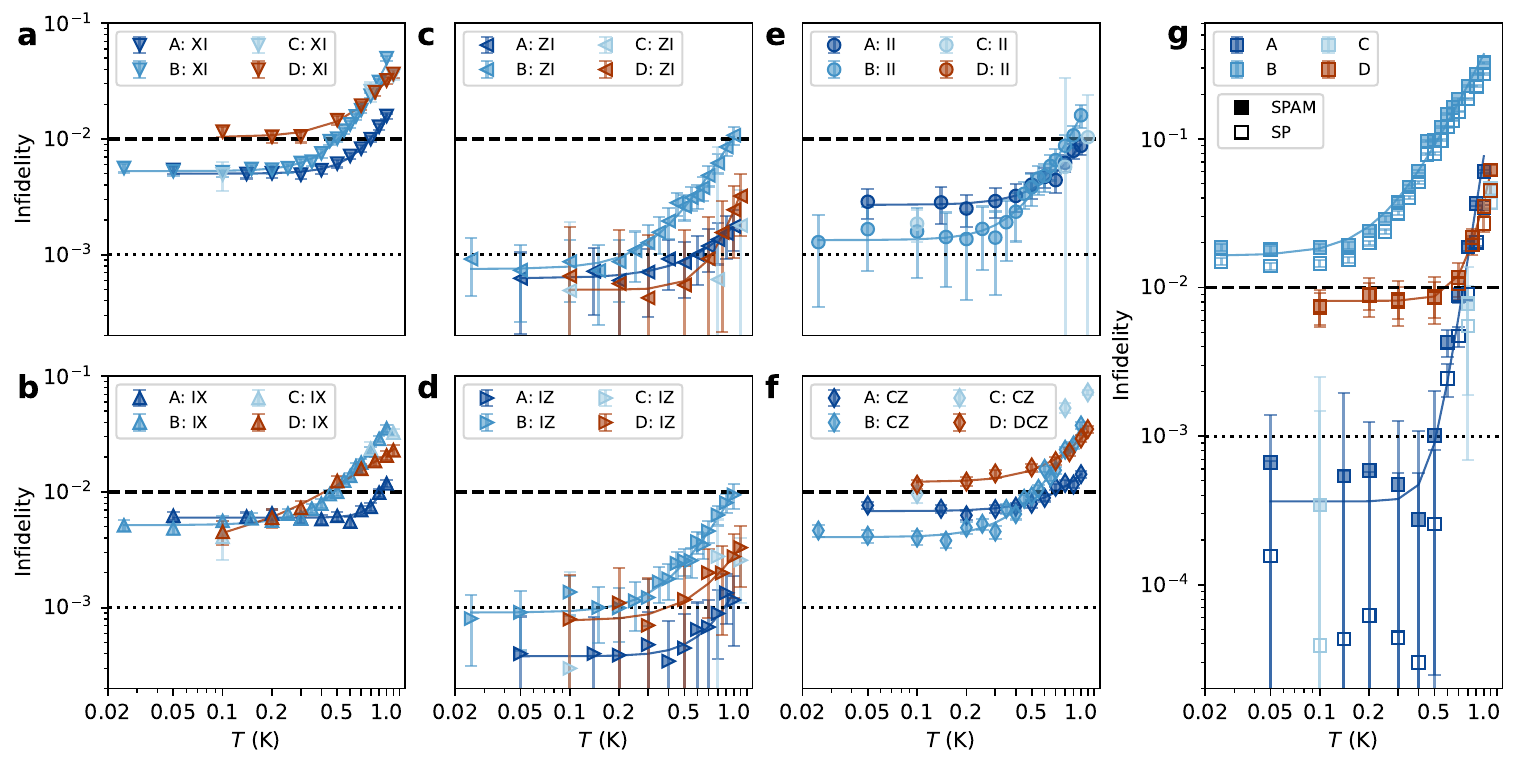}
    \caption{\textbf{Two-qubit system performance.} 
    \textbf{a--g,} Operational infidelity from GST in the two-qubit context as a function of temperature for devices A--D. The solid lines are fits to the power law equation~(\ref{eq:infidelity}). 
    The black dashed (dotted) line represents the \SI{1}{\percent} (\SI{1}{\permille}) infidelity threshold. Error bars represent the \SI{95}{\percent} confidence level.
    }
    \label{fig:main_fig3}
\end{figure*}

\begin{figure}[bth]
    \includegraphics[width = 1\linewidth]{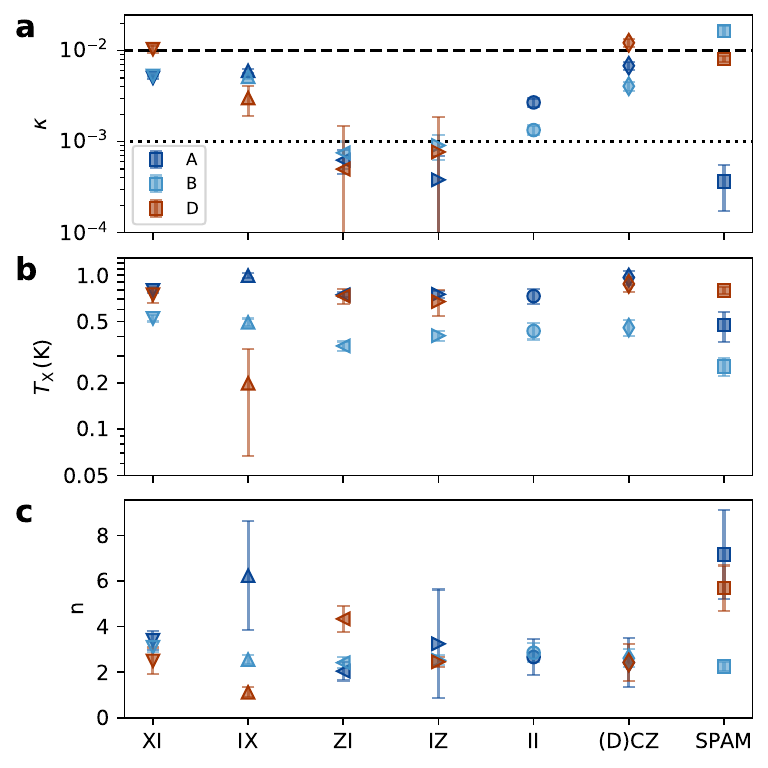}
    \caption{\textbf{Two-qubit system power law fit.} 
    \textbf{a--c,} Power law fitting parameters $\kappa$, $T_{\mathrm{X}}$, and $n$ of equation~\ref{eq:infidelity} to device data A--C in Fig.~\ref{fig:main_fig3}.
    The black dashed (dotted) line represents the \SI{1}{\percent} (\SI{1}{\permille}) infidelity threshold. Error bars represent the \SI{95}{\percent} confidence level.
    }
    \label{fig:main_fig4}
\end{figure}

\noindent
We perform GST experiments on the two-qubit systems comprising the gate sets \{I, XI, IX, ZI, IZ, CZ\} for devices A--C and \{XI, IX, ZI, IZ, DCZ\} for device D. In both cases, we construct a list of ``preparation'' and ``measurement fiducials'' circuits and combine these fiducials with ``germ'' circuits up to length 16, resulting in a list of $\approx$\,13,000 unique sequences~\cite{nielsen_gate_2021}. GST provides a detailed error taxonomy through its decomposition into error generators~\cite{blume-kohout_taxonomy_2022}, enabling insight into the physical origins of operational infidelity. Furthermore, GST generally provides a more stringent assessment of gate performance than other commonly used quantum characterization, verification, and validation (QCVV) methods, which can overestimate gate fidelities~\cite{nielsen_gate_2021,tanttu_assessment_2024}. For details of the gate implementations we refer to previous work on devices A--C~\cite{steinacker_industry-compatible_2025} and device D~\cite{steinacker_bell_2025}. Gate times in devices A--C are $t_{\mathrm{XI}} = t_{\mathrm{IX}} \approx \SI{400}{\nano\second}$ and $t_{\mathrm{I}} = t_{\mathrm{CZ}} \approx \SI{200}{\nano\second}$. For device D, $t_{\mathrm{XI}} = \SI{1200}{\nano\second}$, $t_{\mathrm{IX}} = \SI{568}{\nano\second}$, and $t_{\mathrm{DCZ}} = \SI{4000}{\nano\second}$, with the entangling gate employing dynamical decoupling on both qubits. In all devices, ZI and IZ operations are implemented virtually through microwave phase updates, resulting in gate durations set by the FPGA latency~\cite{steinacker_bell_2025,steinacker_industry-compatible_2025}.

Figure~\ref{fig:main_fig3} summarizes the GST-extracted gate and SPAM infidelities as a function of temperature for devices A--D. At low temperatures, the XI and IX gate infidelity is $\leq 5\cdot 10 ^{-3}$ across all devices at low temperatures (Fig.~\ref{fig:main_fig3}a,b). The larger XI infidelity observed in device D is attributed to its approximately two-fold longer gate duration, which increases the accumulation of dephasing on the idling qubit (Extended Data Figs.~\ref{fig:ExtData_fig1}, \ref{fig:ExtData_fig2}). The virtually implemented ZI and IZ gates exhibit a fidelity of $\leq 10^{-3}$ across all devices at base temperature (compare Fig.~\ref{fig:main_fig3}c,d). The CZ (DCZ) gates show a fidelity of 5--$8 \cdot 10^{-3}$ ($10^{-2}$) (compare Fig.~\ref{fig:main_fig3}f). Devices A and C maintain state preparation (SP) and measurement (SPAM) infidelity at or below $10^{-3}$ ($10^{-2}$) up to \SI{0.5}{\kelvin} (\SI{0.8}{\kelvin}), while device D exhibits SPAM infidelities below $10^{-2}$ between \SI{0.5}{}--\SI{0.7}{\kelvin} (Fig.~\ref{fig:main_fig3}g). Device B shows higher SPAM infidelity owing to less extensive initialization optimization.  
Despite substantial variations in base-temperature performance $\kappa$ across devices (Fig.~\ref{fig:main_fig4}a), the temperature dependence of the GST metrics is remarkably consistent, with most gate and SPAM infidelities exhibiting power-law exponents of $n=2$--4 and crossover temperatures of $T_\mathrm{X}=\SI{0.5}{}$--\SI{0.8}{\kelvin} (Fig.~\ref{fig:main_fig4}b,c). The emergence of a common crossover temperature range across all devices highlights the vicinity of \SI{0.5}{}--\SI{1}{\kelvin} as a particularly relevant regime for silicon spin-qubit operation. The broader implications of this transition are explored through the system-level power analysis below.

Analysis of the GST error generators reveals that operational infidelities across all devices are predominantly driven by stochastic IZ and ZI dephasing errors. As the temperature increases, the reduction in spin coherence and relaxation times amplifies the impact of these idling processes, identifying dephasing as the primary limitation to high-fidelity operation above \SI{1}{\kelvin}. Extending the elevated-temperature operating regime will therefore require further suppression of dephasing through improvements in materials and control, including cleaner oxide interfaces, reduced nuclear-spin concentrations and lower electrical-noise coupling from integrated cryogenic electronics. In addition, continuously driven operating schemes employing dynamical decoupling may provide a pathway towards higher-temperature operation by exploiting the substantially longer rotating-frame coherence times, $T_{2\rho} \gg T_2^*$~\cite{hansen_entangling_2024}.

\section{Power consumption of a quantum computer}
\noindent
Motivated by the observed temperature dependence of operation infidelity, we now model the total power consumption of a utility-scale silicon quantum computer and estimate its optimal operating temperature. Although such a machine will require substantial classical computing resources for qubit control, error decoding, monitoring and compilation, these costs are expected to scale similarly to those of conventional data centers. The quantum-specific power budget is instead dominated by the energy required to maintain cryogenic operation. We therefore focus on the \emph{intensive} energy cost as the key metric, defined as the energy consumed per qubit per gate operation.

\subsection{Quantum error correction requirements}
\noindent
Achieving fault-tolerant operation in a large-scale quantum computer requires a substantial overhead of physical qubits per logical qubit.  As an example, we consider the surface code~\cite{kitaev_fault-tolerant_2003,dennis_topological_2002,raussendorf_fault-tolerant_2007,fowler_surface_2012}. In a typical surface code realization, the logical error rate $p_\text{L}$ scales as~\cite{shaw_quantum_2022}
\begin{equation}
    p_\mathrm{L}\approx \alpha (\beta p)^{\frac{d+1}{2}}, \quad n_\mathrm{q}=2d^2-1
\end{equation}
for $n_\mathrm{q}$ physical qubits, where $p$ is the physical error rate, $d$ is the code distance, while $\alpha$ and $\beta$ are empirical constants determined by error correction code and decoding protocol. For rotated surface codes, $\alpha= 0.3$ and $\beta = 70$~\cite{shaw_quantum_2022}, values that are used throughout this work. To provide a baseline, we assume a system containing 500 logical qubits and calculate the minimum number of physical qubits required to achieve a fixed target logical error rate as a function of operating temperature.

For a fixed logical error rate, increasing the physical error rate $p$ necessitates a larger code distance $d$ must increase,  resulting in a larger number of physical qubits $n_\mathrm{q}$. Consequently, maintaining the same logical computational performance requires greater physical resources and therefore higher power consumption.

To quantify this effect, we estimate the energy required for one surface-code error-correction cycle. For a distance-$d$ rotated surface code, each cycle comprises a well-defined sequence of state-preparation, measurement, single-qubit, two-qubit and idle operations. The corresponding operation counts are summarized in Table~\ref{tab:gatecounts}.
\begin{center}
\begin{table}
\begin{tabular}{ |c|c| } 
 \hline
 Gate type & Gate count  \\\hline\hline 
 State Prep. & $d^2-1$ \\\hline 
 Measurement & $d^2-1$ \\\hline 
 CNOT & $4(d^2-d)$ \\\hline
 Hadamard & $d^2$ \\\hline 
 Idle & $3d^2+4d-4$ \\
 \hline
\end{tabular}
\caption{\label{tab:gatecounts} Operation counts per round of syndrome measurements, broken down by gate type, for a distance-$d$ rotated surface code (excluding logical initialization and readout)~\cite{tomita_surfacecode_2014}.}
\end{table}
\end{center}

\subsection{Power requirements}
\noindent
The power required to operate a fault-tolerant quantum computer can be divided into room-temperature and cryogenic contributions. We refer to the room-temperature contribution as the baseload power, comprising classical control and decoding hardware, operation of the cryogenic plant, generation of global fields and other supporting infrastructure. Although significant, these costs are expected to depend only weakly on both qubit number and operating temperature.
We therefore focus on the cooling power associated with maintaining the quantum processor at a given cryogenic temperature. This contribution is determined by the heat dissipated during qubit operation and scales directly with the number of physical qubits required for fault tolerance. Using the operation counts in Table~\ref{tab:gatecounts} and the energy dissipated by each operation, we estimate the cooling-power requirement as a function of operating temperature, qubit fidelity and the corresponding number of physical qubits required per logical qubit.

Classical control electronics also contribute significantly to the overall power budget. Depending on the architecture, these systems may be located at room temperature, at an intermediate cryogenic stage, or near the qubit operating temperature. Their power consumption can scale approximately linearly with the number of physical qubits, but both its magnitude and temperature dependence are strongly architecture specific. To retain generality, we therefore omit this contribution from the present analysis and focus on the cooling power associated with the quantum processor itself. This simplification does not alter the central conclusions regarding the existence of an optimal operating temperature.

The total power consumption is therefore modeled as
\begin{eqnarray}
    P_{\rm{total}} & = & P_{\rm{base}} + P_{\rm{control}} \\
    & & + P_{\rm{global}} + N_{\rm{logical}}d^2 (P_{\rm{SPAM}} + P_{\rm{2Q}} + P_{\rm{1Q}}), \nonumber
\end{eqnarray}
  
where $P_{\rm{base}}$ is the baseload power required to operate the overall system, $P_{\rm{control}}$ is the power associated with classical control and decoding hardware, and $P_{\rm{global}}$ accounts for cryogenic power overheads that are independent of qubit number but arise from operation of the quantum processor. The terms $P_{\rm{SPAM}}$, $P_{\rm{2Q}}$, and $P_{\rm{1Q}}$ denote the power dissipated by state preparation and measurement, two-qubit gates and single-qubit gates, respectively. These contributions scale with the number of logical qubits, $N_{\rm{logical}}$, and distance squared $d^2$, reflecting the increasing physical-qubit overhead required to maintain a fixed logical error rate as the physical error rate increases.

In principle, an additional contribution $P_{\rm{couple}}$ should be included to account for long-range communication between qubits, such as spin-to-photon conversion~\cite{dobinson_electrically_2025, methodstransverselongitudinalspinphoton_2023} or coherent shuttling~\cite{yoneda_coherent_2021, 3dot_shuttling_2026}. Although this contribution may be non-negligible in a practical implementation, it scales approximately linearly with $d$, whereas the dominant power contributions scale with $d^2$. Consequently, it does not significantly affect the conclusions of the present analysis (Extended Data Fig.~\ref{fig:ExtData_fig3}).

For silicon-qubit research, cryogenic operation is typically achieved using dilution refrigerators. At utility scale, however, the cooling technology and operating temperature become system-level optimization parameters. To explore the resulting trade-offs without assuming a particular cryogenic architecture, we model the refrigeration system using a generalized Carnot refrigerator, 
\begin{equation}
    P = \frac{\dot{Q}}{\eta}\frac{(T_{\rm{H}} - T_{\rm{C}})}{T_{\rm{C}}}, 
\end{equation}
where $\dot{Q}$ is the heat dissipated at the qubit layer, $P$ is the electrical power required to remove this heat, $T_{\rm{H}} = \SI{297}{\kelvin}$ is the ambient temperature, $T_{\rm{C}} = T_{\rm{q}}$ is the qubit operating temperature and $\eta=0.25$ is the refrigerator efficiency relative to the Carnot limit, consistent with values reported for large-scale cryogenic systems (\SI{28.5}{\percent})~\cite{cern_efficiency_2012}. Although practical implementations will depend on the cooling technology and overall system architecture, the dominant thermodynamic scaling is captured by this generalized Carnot model.

\subsection{Silicon quantum computer as an example}
\noindent
To provide concrete estimates, we now consider a specific architecture based on spin qubits in gate-defined silicon quantum dots fabricated using industrial foundry processes. We assume qubit performance comparable to, or slightly exceeding, that measured in the previous section. Several control modalities are compatible with this platform, including direct electron-spin-resonance (ESR) control~\cite{loss_quantum_1998}, microwave-dressed qubits~\cite{laucht_dressed_2017,seedhouse_quantum_2021}, exchange-only qubits~\cite{divincenzo_universal_2000,eng_isotopically_2015,andrews_quantifying_2019,weinstein_universal_2023}, and SMART qubits~\cite{hansen_pulse_2021,hansen_implementation_2022,hansen_entangling_2024}. Here we focus on SMART qubits, which combine a globally applied microwave field in the GHz range with local control tones in the MHz range. This approach reduces the number of local high-frequency control lines required at the qubit layer, providing a favorable trade-off between cryogenic heat dissipation, control complexity and quantum-dot footprint.

To estimate the cooling power for this architecture, the operation counts in Table~\ref{tab:gatecounts} must first be expressed in the native SMART gate set. CNOT gates are not native to this architecture and are instead implemented using two $\sqrt{\rm{SWAP}}$ gates together with four single-qubit rotations~\cite{seedhouse_quantum_2021}. This decomposition increases the number of physical operations required per error-correction cycle and is included explicitly in the power estimates presented below. Because the compilation is not optimized for the native gate set, these estimates should be regarded as a conservative upper bound.

At the frequencies and temperatures relevant to silicon spin qubits, dielectric losses associated with the MHz control voltages applied to the quantum-dot gate electrodes are expected to dominate the local heat dissipation during qubit operations. The corresponding dissipated power can be approximated as $\dot{Q}=CV^2 f D$ where $C\approx \SI{3}{\femto \farad}$ is the gate capacitance, $V\approx\SI{0.8}{\volt}$ is the applied voltage, $f \approx \SI{1}{\mega \hertz}$ is the local modulation frequency, and $D=1$ is the assumed duty cycle.
The SMART architecture additionally requires a globally applied GHz microwave field to dress the qubits, which contributes a qubit-independent cryogenic heat load. Unless otherwise stated, we assume $\dot{Q}_{\rm{global}}=\SI{1}{\milli \watt}$.

to model the temperature dependence of the error-correction overhead, we assume identical physical error rates for all operations, such that $p_{\rm{SPAM}}=p_{\rm{2Q}}=p_{\rm{1Q}}=p_{\rm{idle}} = p = 1-\cal{F}$. The physical error rate is therefore taken directly from the measured operation infidelity (Eq.~\ref{eq:infidelity}), using $\kappa=0.0012$, $T_{\mathrm{X}}=\SI{1.14}{\kelvin}$ and $n=3.41$ as extracted from device E~\cite{huang_high-fidelity_2024} and shown in Fig.~\ref{fig:main_fig0}a. Although the fidelities reported for SMART qubits~\cite{seedhouse_quantum_2021,hansen_pulse_2021,hansen_implementation_2022,hansen_entangling_2024} are expected to be comparable to or better than the bare qubit fidelities measured here, we conservatively assume the same error rate for all operations. This provides an upper bound on the resulting error-correction overhead. We further assume $n_{\mathrm{L}} = 500$ logical qubits and calculate the number of physical qubits required to achieve a fixed target logical error rate as a function of the operating temperature $T_{\mathrm{q}}$.

\begin{figure} 
    \includegraphics[width = 1\linewidth]{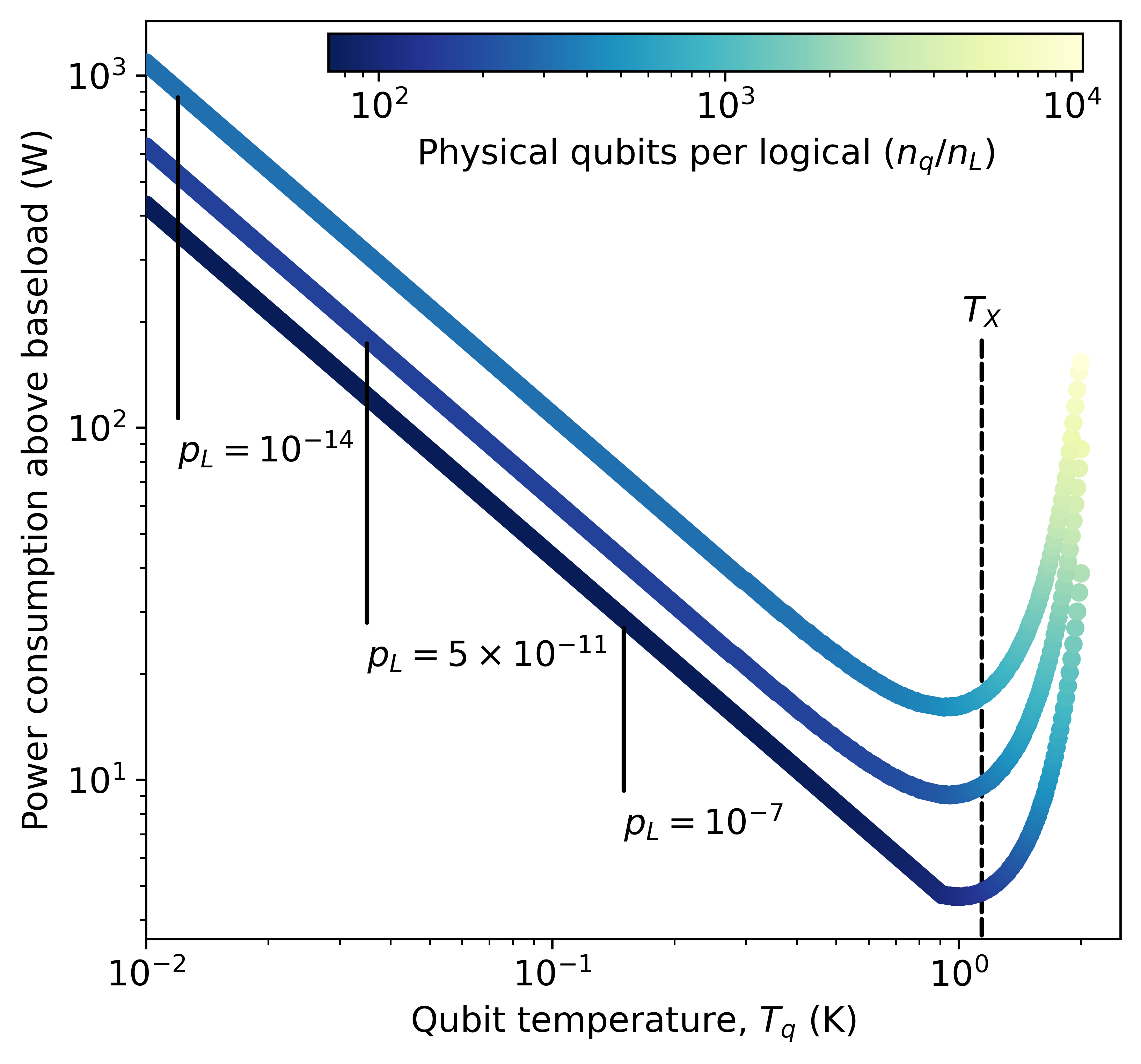}
    \caption{\textbf{Power consumption as a function of operating temperature.} 
    Power consumption for three target logical error rates $p_L$, assuming a fixed number of logical qubits ($n_\text{L}=500$). A finite optimum emerges from the trade-off between cooling requirements and quantum-error-correction overhead.}
    \label{fig:main_fig5}
\end{figure}

The resulting contours of constant logical error rate are shown in Fig.~\ref{fig:main_fig5}. A central result of this work is that the minimum-power operating point of a fault-tolerant quantum computer does not, in general, coincide with the temperature at which the qubits exhibit their highest fidelity. Instead, the model predicts a distinct optimum at finite temperature, arising from the competition between refrigeration cost and quantum-error-correction overhead. Increasing the operating temperature reduces the electrical power required for cooling, but also increases the physical error rate, requiring more physical qubits to realize a fixed number of logical qubits. The additional cooling power associated with this growing physical-qubit overhead ultimately outweighs the thermodynamic benefits of operating at higher temperature, producing a well-defined minimum in the total power consumption. Because this mechanism depends only on the relationship between cooling requirements and error-correction overheads, it is expected to be a generic feature of cryogenic quantum-computing architectures. Alternative parameterizations and fidelity models presented in the Extended Data produce qualitatively similar behavior, supporting the generality of this conclusion.

For silicon quantum-dot spin qubits exhibiting performance comparable to the devices studied here, the model predicts an optimal operating temperature in the range of approximately \SI{0.5}{\kelvin}--\SI{1.5}{\kelvin}. Improved understanding of the physical mechanisms underlying the temperature dependence of gate infidelity will enable advances in materials, device design and control techniques that can further reduce power consumption and shift the optimum operating point (Extended Data Fig.~\ref{fig:ExtData_fig4}). More generally, our results show that the operating temperature of a quantum computer should be optimized at the system level, balancing refrigeration requirements against quantum-error-correction overhead. As quantum processors scale from laboratory demonstrators to commercially relevant machines comprising thousands to millions of physical qubits, this trade-off will become an increasingly important determinant of both operating cost and system scale.

\section{Discussion}
\noindent
Our results show that the operating temperature of a quantum computer should be treated as a system-level design parameter rather than simply a device-level performance metric. The existence of a finite optimal operating temperature highlights the importance of jointly optimizing qubit fidelity, error-correction overhead and refrigeration requirements when designing utility-scale quantum computers. In this context, the demonstration of high-fidelity qubit operation near \SI{1}{\kelvin} in industrially fabricated silicon processors is particularly significant, as it opens a pathway towards integrating classical control electronics and qubits within a common cryogenic platform while maintaining favorable overall system efficiency.

Although the industrial devices benefit from an optimized low charge-noise environment~\cite{stuyck_integrated_2020,li_flexible_2020,stuyck_uniform_2021,elsayed_low_2024,steinacker_industry-compatible_2025}, the temperature dependence of the infidelity follows a similar trend to that observed in the academic devices (D and F), indicating that the mechanisms limiting elevated-temperature operation are not eliminated by industrial fabrication alone. At the same time, device E, fabricated on an isotopically enriched silicon substrate containing only \SI{50}{ppm} residual $^{29}$Si, maintains superior performance at $T\geq\SI{1}{\kelvin}$~\cite{huang_high-fidelity_2024}. This observation suggests that residual nuclear-spin noise remains an important contributor to fidelity degradation at elevated temperatures. Together, these results indicate that further advances in both isotopic purification and device engineering will likely be required to achieve the fidelities necessary for large-scale fault-tolerant quantum computing.

The predictive capability of the framework presented here will improve as larger qubit populations become available for statistical analysis. Achieving this will require automated tuning and characterization workflows capable of operating across large device arrays~\cite{zwolak_colloquium_2023}. SiMOS qubits fabricated in an industrial foundry environment have already demonstrated a high degree of reproducibility~\cite{steinacker_industry-compatible_2025}, making them a promising platform for such large-scale studies. Equally important is the development of a deeper understanding of the physical mechanisms responsible for the observed temperature-dependent degradation in fidelity. Identifying and mitigating these noise sources will enable further improvements in qubit performance, potentially shifting the optimal operating temperature and reducing the power requirements of fault-tolerant quantum computation. Ultimately, our results show that optimizing a quantum computer requires consideration of both device physics and system-level resource costs, linking qubit fidelity, error-correction overhead and refrigeration requirements within a common design framework.

\bibliography{mybib}

\section*{Acknowledgments}
\noindent We acknowledge support from the Australian Research Council (FL190100167) and the U.S. Army Research Office (W911NF-23-10092). P. S. and A. N. acknowledge support from the Sydney Quantum Academy. P. S. acknowledges support from the Baxter Charitable Foundation. N. D. S. and K. W. C are the recipients of an Australian Research Council Industrial Fellowship (project numbers IE240100252 and IM230100396) funded by the Australian Government. A. E. S. is suppported by the Singapore National Research Foundation (NRF) and French National Research Agency (ANR) joint project “QuRes” (Grant No. ANR-21-CE47-0019; NRF2021-NRF-ANR005), she also acknowledges the support of the “OECQ project” (Contract DOS0226235/00) financed by the French state (via France 2030) and Next Generation EU (via France Relance).

\section*{Author Contributions}
\noindent W. H. L., K. W. C., F. E. H., C. C. E. and N. D. S. designed the devices.
P. S. conducted the experiments with supervision from T. T. and N. D. S. as well as input from C. H. Y., A. L., A. S. D., An. S., J. H. C..
S. S., E. V., S. K. B., and A. N. assisted with the experimental setup.
S. P., B. H.-J., and J. P. D. assisted with measurements of device B.
S. S. assisted with measurements of device C.
N. D. S. wrote the GST experiment implementation with input from M. F.. 
M. F. helped with the GST analysis. 
A. E. S., P. Y. M., and Al. S. developed the power model and QEC implications with input from An. S. and J. H. C..
P. S. and J. H. C. wrote the manuscript, with input from all authors.
W. H. L., A. L., A. S. D., and A. S. supervised the project. 

\section*{Competing Interests}
\noindent A. S. D. is CEO and a director of Diraq Pty Ltd. N. D. S, T. T., M. K. F., S. S., E. V., S. K. B., P. Y. M., Al. S., F. E. H., K. W. C., C. C. E., C. H. Y., W. H. L., A. L., A. S. D., and An. S. declare equity interest in Diraq. Other authors declare no competing interest.


\section{Extended Material}
\setcounter{figure}{0}
\setcounter{table}{0}
\captionsetup[figure]{name={\bf{Extended Data Fig.}},labelsep=line,justification=raggedright,font=small,singlelinecheck=false}

\subsection{Influence of temperature dependence functional form}\label{sec:ExtendedFunctionalForm}
\noindent
A number of functional forms for the temperature dependence of qubit gate fidelity are plausible. Throughout the main text we use a power-law function with a constant (temperature independent) term of the form
\begin{equation}\label{eq:infidelity_func}
    {1-\cal{F}} = \kappa \left[ 1 + \left( \frac{T}{T_X} \right)^n \right],
\end{equation}
as this can be easily mapped to standard models of charge noise, spin noise and electron-phonon coupling processes.

Other models the could be employed include a temperature dependence based on a bosonic bath,
\begin{equation}\label{eq:infidelity_bosonic}
    {1-\cal{F}} = \kappa_0 + \frac{\kappa_1}{\exp(\kappa_2/T) - 1},
\end{equation}
as one would expect if coupling directly to phonons, or an exponential dependence of the type you would expect for quasiparticle excitations,
\begin{equation}\label{eq:infidelity_phonon}
    {1-\cal{F}} = \kappa_0 + \kappa_1 \exp(-\kappa_2/T).
\end{equation}

As can be seen in Fig.~\ref{fig:main_fig0}a, all three functional forms can be used to fit the data. The only noticeable difference appears for $T_q >\SI{1.4}{\kelvin}$. As a result the optimal operating temperature is largely unchanged as above $\SI{1}{\kelvin}$ the power is dominated by the additional cost required by the higher error correction requirements. The reduced infidelity at higher temperatures (for the alternative models) results in a reduced code distance and lower power. However both still ultimately increase with temperature, resulting in a similar minima in total power. 

\subsection{Breakdown of contributions to two-qubit system infidelity}\label{sec:ExtendedInfidelityContributions}
\begin{figure*}[tb]
    \includegraphics[width=1\textwidth,angle = 0]{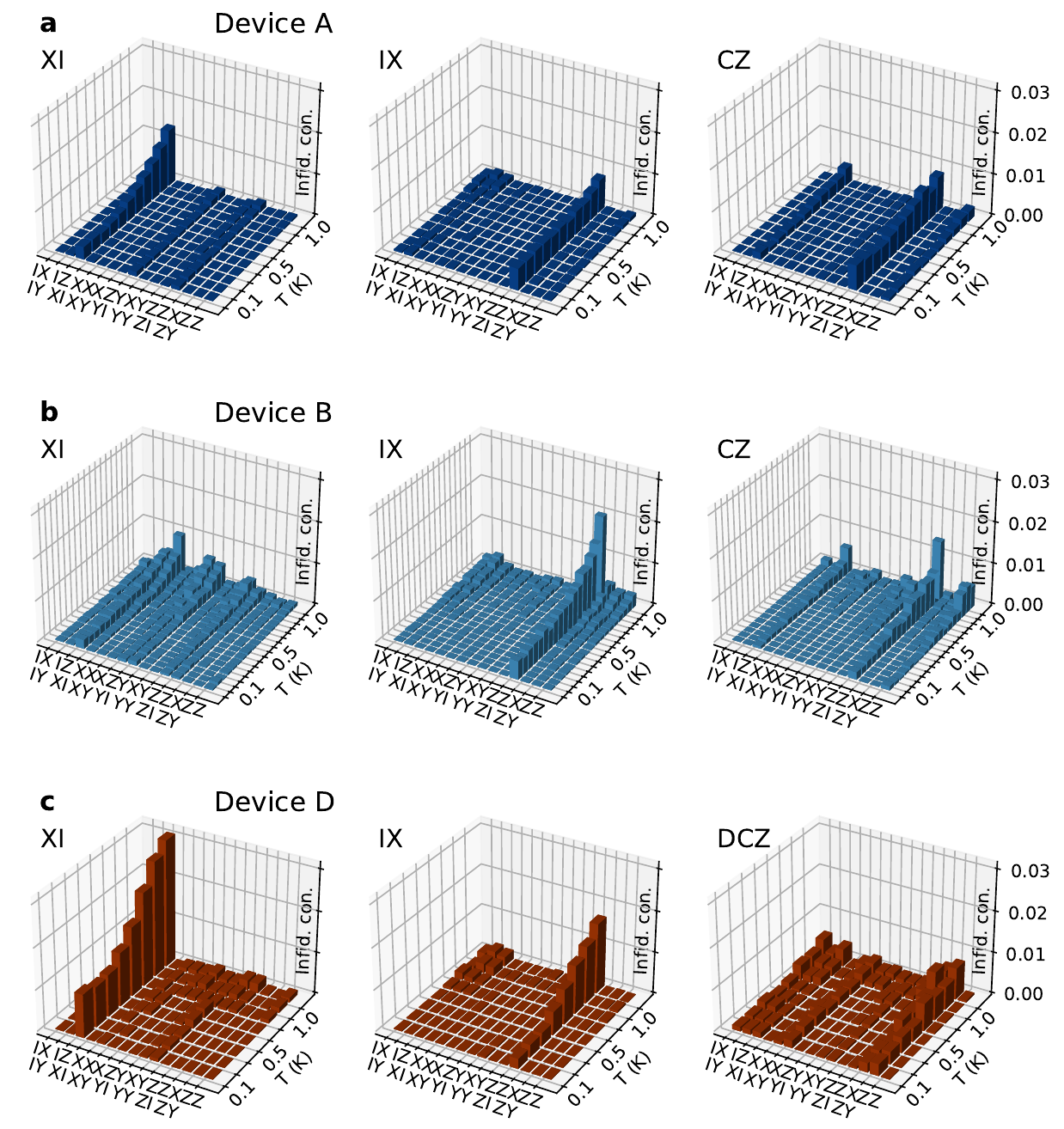}
    \caption{\textbf{Two-qubit system stochastic infidelity contribution.} 
    \textbf{a--c}, Breakdown of the stochastic infidelity contributions to the XI, IX, and CZ/DCZ gate as a function fo temperature $T$ for device A (\textbf{a}), B (\textbf{b}), and D (\textbf{c}). Dephasing error generators (stochastic IZ and ZI) are dominating the infidelity for all gates but the DCZ gate, particularly at elevated temperatures of up to $\SI{1.1}{\kelvin}$. The DCZ gate mitigates dephasing noise due to its intrinsic dynamical decoupling sequence.
    }
    \label{fig:ExtData_fig1}
\end{figure*}

\begin{figure*}[tb]
    \includegraphics[width=1\textwidth,angle = 0]{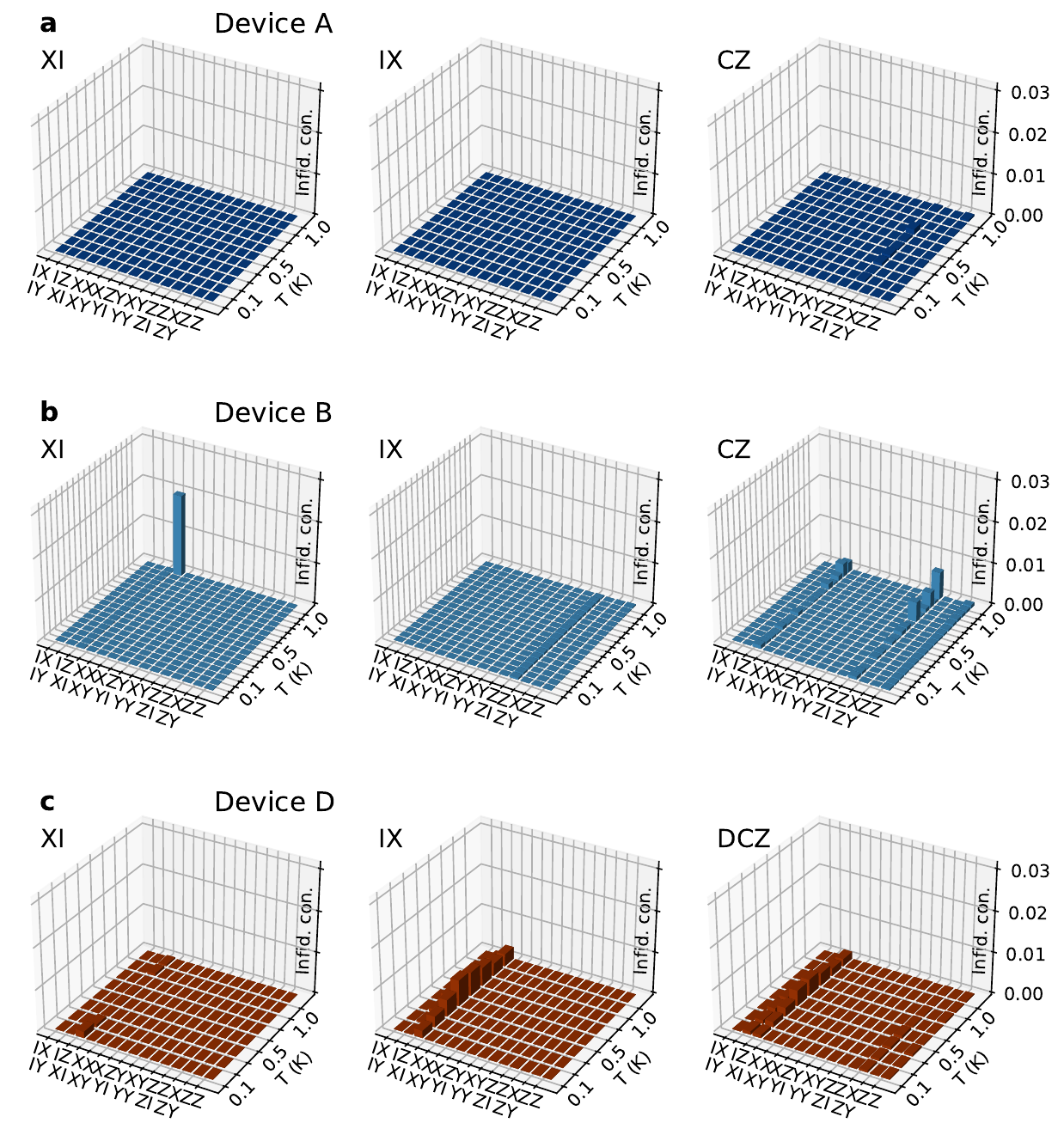}
    \caption{\textbf{Two-qubit system Hamiltonian infidelity contribution.}
    \textbf{a--c},  Breakdown of the Hamiltonian infidelity contributions to the XI, IX, and CZ/DCZ gate as a function of temperature $T$ for device A (\textbf{a}), B (\textbf{b}), and D (\textbf{c}). The DCZ gate shows similar Hamiltonian infidelity contributions as the single-qubit gates in the two-qubit context it is constructed of to implement the dynamical decoupling sequence.
}
    \label{fig:ExtData_fig2}
\end{figure*}

\noindent
The relative stochastic (Extended Data Fig.~\ref{fig:ExtData_fig1}) and Hamiltonian (Extended Data Fig.~\ref{fig:ExtData_fig2}) contributions to the gate infidelity as a function of temperature, showing the dominate error modes limiting the performance of the devices studied in the main text.

\subsection{Power consumption due to shuttling}\label{sec:ExtendedPowerShuttling}
\begin{figure}[tb]
    \includegraphics[width = 1\linewidth]{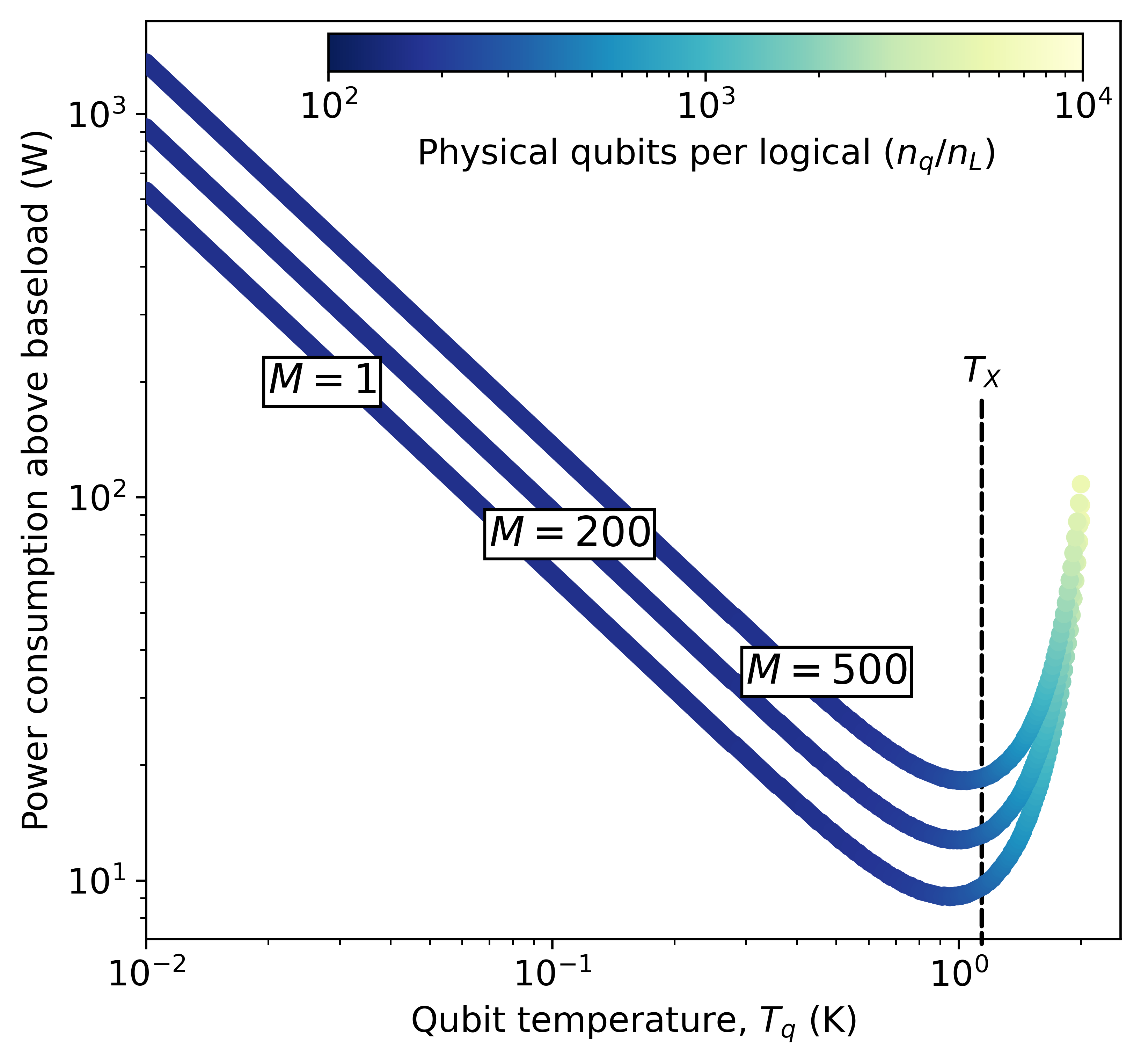}
    \caption{\textbf{Power consumption as a function of operating temperature.} 
    Power consumption for $p_\text{L}=5\times10^{-11}$ and $n_\text{L}=500$, for $M=$1, 200, 500, as a function of temperature. This illustrates that the influence of this model of shuttling cost on total power is only appreciable at low temperatures and becomes negligible as the code distance increases (due to the quadratic dependence on $d$ of the surface code circuits dominating over the linear dependence on $d$ of the shuttling).}
    \label{fig:ExtData_fig3}
\end{figure}

\noindent
Although the power consumption of long distance links is extremely dependent on the technology being employed, it is still useful to show the result of including a simplistic model. Here we assume that long distance qubit transport is performed via shuttling~\cite{yoneda_coherent_2021, 3dot_shuttling_2026}. As an initial model, we assume the power scales linearly with distance shuttled ($\sim d$) and that the power cost is equivalent to $M$ idle gate operations, \textit{ie.}, $P_{\rm{couple}}=N_{\rm{logical}} M  P_{\mathrm{1Q}} \times d$. Extended Data Fig.~\ref{fig:ExtData_fig3} shows the effect on the power consumption curve for $p_\text{L}=5\times10^{-11}$ as a function of $M$.

\subsection{Dependence of power consumption on temperature dependent fidelity performance}\label{sec:ExtendedPowerFidelity}
\begin{figure}[tb] 
    \includegraphics[width = 1\linewidth]{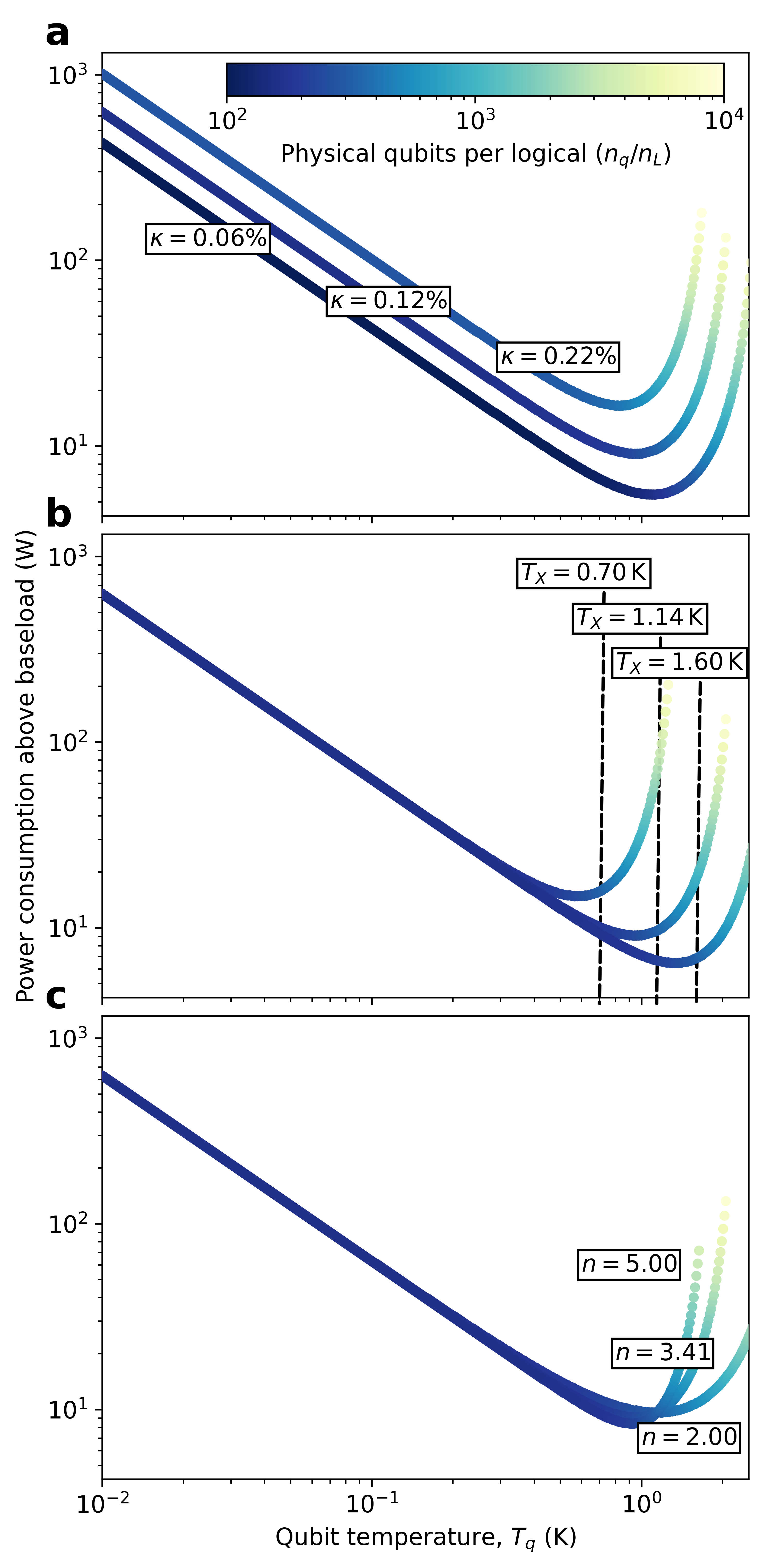}
    \caption{\textbf{Power consumption as a function of operating temperature.} 
    Power consumption for various characteristics of the power-law temperature response, for $p_\text{L}=5\times10^{-11}$ and $n_\text{L}=500$. The influence of varying the a) low temperature infidelity $\kappa$, b) cross-over temperature $T_X$ and c) temperature exponent $n$ parameters are illustrated with examples that span the fitting values.}
    \label{fig:ExtData_fig4}
\end{figure}

\noindent
In the presented data there is a clear temperature independent contribution to the infidelity. This kind of low temperature asymptote for the infidelity will differ greatly from architecture to architecture, and even from design to design in the same architecture. The interesting observation is that due to the strong power dependence of cooling on the operating temperature, the optimal operating temperature will still always appear at a finite temperature. Therefore improvements in the low temperature noise properties translate to performance improvements and reduced power consumption, but don't necessarily reduce the optimal operating temperature significantly, as can be see in Extended Data Fig.~\ref{fig:ExtData_fig4}a.

Similarly, significant changes in the temperature dependent exponent $n$ do not have a significant effect as for $T_\text{q}>T_\text{X}$ the error correction cost quickly dominates the power cost, Extended Data Fig.~\ref{fig:ExtData_fig4}c. However for small $n$, there is less variation in power consumed close to the optimal point, suggestion high stability to temperature variation.

The cross-over temperature $T_\text{X}$ is more interesting as this dominates the trade-off between cryogenic and error correct power costs, and effectively sets the optimal operating temperature, Extended Data Fig.~\ref{fig:ExtData_fig4}b. This implies that methods that increase $T_\text{X}$ will have a large effect on the overall operating cost of large scale systems.

\end{document}